# Innovation and Evolution of Business Relations and Networks: Theory and Method

**Ian F. Wilkinson***

**Louise C. Young****

*) Discipline of Marketing, The University of Sydney and Department of Entrepreneurship and Relationship Management, University of Southern Denmark
**) School of Business, University of Western Sydney and Department of Entrepreneurship and Relationship Management, University of Southern Denmark

**ABSTRACT**

We introduce a complex systems perspective on innovation in networks in which innovation is conceptualized as a form of creative act associated with the dynamics and evolution of business network. We show how innovation is a form of creative act that involves the creation of new ideas and their exploitation, in which new ideas come from combining and recombining existing ideas in new ways that have value. We stress the need to move away from traditional linear, comparative static variables based theories and models to more nonlinear, dynamic, evolutionary mechanism and process based theories models of business networks. This calls for different types of methods including systematic case histories and agent-based models

## 1. Introduction

Innovation in business networks is about how and why things change over time; about the dynamics and evolution of products and services and business organisations, including relations and networks themselves (Wilkinson, 2008). In this paper we argue that the concept of innovation hides a deeper and more profound logic of change and change processes. The concept of innovation implicitly suggests that non-change is the norm. But is this true? Things are changing and evolving all the time in markets; they are not operating in equilibrium but far from it. Products and services are continually being produced, consumed, bought and sold. This may create loyalty and non change in a specific sphere but this in itself is likely to influence changes in other spheres (i.e. non change can be the cause change somewhere else). Transactions of different kinds are being brought about with the same or different transaction partners, relations and networks are continually being formed and reformed. Considerable effort is often involved to keep reasonable amounts of stability within relationships and networks in a world of change. Business life is one continuous flux or flow of action, reaction and adaptation over time.

Some patterns of behaviour are repeated or reproduced over time; habits of behaviour and thought develop. Firms may continue to interact with each other for a long time – as we have seen in previous IMP case studies. The same basic shape of a business network may also continue to exist over time as people and firms get to know each other, trust each other and become loyal and committed to dealing with particular customers and/or suppliers. This means that they try to keep interacting with the same partners, even though things are continually changing at other levels. The personnel involved change, knowledge and experience grows, markets change, new opportunities emerge, technologies change, information and ideas move about in the network, changes in other relations impact on a focal relation. And so business life proceeds. Change, change, change, all the time. Business life, indeed all of life is not physics. Firms are not billiard balls subject to rules of behaviour like conservation of momentum. We make up our rules of behaviour, within biological and technical limits, and change and evolve them. Business life does not end it goes on forever and has a long history.

And the pace of change is increasing. Lifecycles of product acceptance are shorter than they were 20 years and the rate of product introduction is also increasing. So why do we “privilege” change rather than non-change? Surely it is becoming more and more difficult to keep reproducing past patterns of behaviour. It is change rather than non-change that is the norm.

Changing is a fundamental biological and psychological process. We change to grow, to develop and to survive. At an individual level, we learn and build on what we know. In business we react and adapt to changing conditions in the market and environment. There is a basic business axiom, “change or die”. To be sure, there are forces that work to countervail change - inertia and resistance to change occur as people and firms try to stick to old habits, engage in selective and self serving perceptions and are subject to the “not invented here” syndrome (Rong and Wilkinson, 2011). Lack of action is one part of a network influence other parts and in itself is likely to produce change elsewhere.

Over time social and business systems are continually being reproduced, or not. Reproduction is an active process of reconstructing and redoing things – social and business systems are not like billiard balls subject to the conservation of momentum law that says an object will continue in the same direction unless some external force changes it! Unlike billiard balls, social and business systems are complex adaptive systems (CAS). As such, properties of CAS describe these systems’ properties. These systems show properties of emergence, where collective and group behaviour is different from individual behaviour; nonlinearity, i.e. the behaviour of the whole is more than the sum of the behaviour of its parts. The future cannot be extrapoloated from past, there are webs or networks of interaction rather than evenly distributed interaction, and the degree of diversity of an organisation and its network are important to its survival generally equates with greater fitness (performance) and greater opportunity. The diversity of the organisation must match that of its environment in order to survive, as Ashby first encapsulated in his principle of requisite variety. All of these properties are conducive to change as systems need to constantly adjust to the unanticipated which will be less likely to lead to an unchanged state and more likely to lead to a changed state.

Here we argue that innovation is part of a much larger process of change and evolution that characterizes all complex living systems – biological, economic, business, social and cultural. And there are similarities in the way change and evolution occur across all these types of systems. This is in contrast to a traditional focus on innovation in business which represents a comparative static, actor-centered perspective on business and life, derived from seeing living systems as similar to physical and mechanical systems in which there is conservation of momentum.

Societies and business systems and networks are not like a gas. Social science is not physics. Both comprise numerous interacting entities: gases are made up of randomly interacting, non-evolving, inanimate, homogenous atoms and molecules. In contrast, societies and business systems are made up of networks of interacting, evolving, heterogenous, sentient people and organisations. You cannot represent social and business systems in terms of average individual behaviour or mean field approximations, as many economic models tend to do. The systems are far too complex, non linear, nuanced and dynamic, even though some of the stylized emergent general facts, such as power law distributions of extinctions of firms financial markets (Ciarella et al), duration of relations and duration of superior performance of firms in industries(Wiggins, 2002) observed about such complex systems can be approximated with physics type models(Cook, 2003, Guilmi, 2004, Ormerod and Rosewell, 2009, Ormerod, 2006) (Wiggins, 2002, Kauffman, 1992). For the individual people and firms and their relations and networks it is constant change all the way down. Imagine trying to approximate the behaviour of animals by having a representative animal – would a particular fish do – with some random environment to move about in and feed and have babies? Or, suppose we used lots of random animals that are homogeneous or, better, make them have some degree of heterogeneity in terms of their fitness (whatever that means) with an average degree of interconnectivity to represent the Earth's ecological system. If I did that you would laugh at me. But this is essentially what macro economists and financial market modelers do and the way they think about those systems. How much can that help us tell an individual firm or manager or animal how they do and should behave; would you trust government policy or environment management systems designed based on such models.

In complex adaptive social systems, and business networks, large scale order emerges in bottom-up self-organising ways that are not yet well understood. But the science of complex systems is revolutionizing theory and method in all sciences. It is moving us away from a linear, comparative static, variables-based view of the world to a view that is non-linear, dynamic, process and mechanism based (Jorg, 2011). Such a view, we believe, can help us to better understand, model and influence innovation in networks.

In this view, innovation is a process by which new ways of behaving occur; it is a form of evolution. Innovation involves two kinds of process: a) the emergence or evolution of new ideas, which may also be described as entrepreneurial or creative acts or opportunity recognition; and b) the development and exploitation of these new ideas. These two processes are distinct but interconnected, they feed off each other (Chandra, 2009, Chandra, 2012, Styles, forthcoming). The processes of development and exploitation of discovered opportunities in turn lead to the discovery of new ones.

## 2. The Evolution of New Ideas.

New ideas have history. They arise from the combining and recombining of existing ideas in new ways that have some value. Knowledge builds on itself. The behaviour and organization of a social system or business network reflects the culture underlying it - the set of ideas characterising the actors involved. Cultural evolution refers to the evolution of ideas and the movement of those changed ideas through a connected social system. This is essentially a creative act and, as Arthur Koestler has pointed out, the creative process is essentially the same in art, science and humour.

Earlier theories of innovation and creativity tended to focus on the characteristics of individual actors in trying to explain creativity in all its forms, including innovation, entrepreurship, scientific progress, invention and technological change. Creative people and organisations have different characteristics to less or non-creative ones. But this “lone genius” type explanation is rather inadequate, as (Ebel, 1974) first pointed out. It is a description of behaviour masquerading as an explanation - people and firms act creatively because they are creative. Individual differences do matter but it is not as simple as identifying magic ingredients or orientations. What matters also is how individual are positioned in the flow of ideas among people and organisations over time and space – their location in socio-economic networks – as well as how open they are to absorbing new ideas and being motivated and thus able to combine and recombine the ideas they have access to in ways that have value:

> “Entrepreneurs and inventors are no smarter, no more courageous, tenacious, or rebellious than the rest of us – they are simply better connected” (Hargadon, 2003)

The act of creation in all its forms is a psychological act of recombining ideas (and other things) or what Matt Ridley refers to as “ideas having sex with each other” (Ridley, 2011). The sexual encounters of ideas that are possible depend on the pool of ideas that are in play at any time and place and, more specifically what are known to particular people and firms. This depends on those entities’ networks and history.

### 2.1 The Role of Relations and Networks

The role of relations and networks, both business and social, in the innovation process cannot be underestimated. But they play two quite different roles in the discovery and creation of new opportunities and ideas and in their exploitation.  The distinction is captured well in Podolny's concept of networks as prisms and pipes (Podolny, 2001). As prisms, relations and networks are the means by which firms extend their eyes, ears and other senses.  They are the means by which knowledge and ideas move around and people and firms encounter new knowledge and ideas. This can be both through deliberate search with and through others and through more serendipitous processes. Either way, the "self" of the firm is extended beyond its own boundaries, resulting in the enhancement of its adaptive properties (Wilkinson and Young 2005).  Considering these in complexity terms, there is a group comprised of people, firms and other organizations that functions differently than does any firm or organization comprising it, which has properties not divisible or attributable to particular actors in the system (i.e. the interactions of firms have properties distinct from the firms themselves) and these properties can and do result in greater fitness for the firms within the network as well as for the network itself and is selected for in the process of evolution (Ladley et al., 2011, Henrich, 2004).

Through these means, new ideas come to the minds of people to be combined and recombined and confronted with other ideas, leading to acts of creation and innovation. Elsewhere we have discussed the specifics of the "soft assembling" that characterize the building of new ideas and particularly strategies (Clark in Wilkinson and Young 2005), here we focus on the output of these processes. These processes continue over time and when combined with outcomes and experiences of an actor's own direct actions and interactions serve to shape what a person and firm knows and does not know at any time and place.  In other words, history matters in determining the set of ideas a person or firm has at any time and place and hence the creative acts they are (then) capable of.

This is referred to as their "prior knowledge" (Shane, 2000), which shapes what a person or firm can and cannot see and how they respond to incoming information, including new ideas. Their motivations, other skills and capabilities and the context in which they operate are path dependent i.e. come about from the unique historical path in which they have participated which opened future possibilities and closed off others.  This past (which cannot be changed, though it can be reinterpreted) determines which, if any, of these creative acts occur and what can and cannot be done about them. These processes lie at the heart of the theories of knowledge and entrepreneurship espoused by Schumpeter (Schumpeter, 1934), Hayek (Hayek, 1945) and further developed later by  Kirzner (Kirzner, 1973, Kirzner, 1997)

The role of relations and networks in shaping innovation is well illustrated in case studies of opportunity recognition in international ventures (Chandra, 2009, Chandra, 2012).  The first example concerns the discovery of an international marketing opportunity in China for a T-Shirt designer, who was looking for a textile manufacturer in China. As the owner explains:

> "I approached a friend in Melbourne who was a pattern maker at a big Australian design company…. The pattern maker introduced me to a Guangzhou factory owner…. During our first summer design launch, the owner of the Guangzhou factory and her daughter also came to Australia to see this presentation and were really impressed. She said that she would be interested to develop 2Spot together in China and that it would sell very well because the market was huge and they loved things European. It sort of fell into place because the Chinese company approached us and said you should open a store in China and do you want to be a partner in it?" (Chandra, 2012)

Here the T-Shirt designer seeking a producer in China uses his personal networks to contact someone in the industry who uses their networks to recommend a potential supplier. The Chinese supplier sees the opportunity in China and persuades the designer to come on board.

Here is another example:

> "We are quite respected in the industry for high quality products….we have served architectural firms in Australia well… one of them is Thompson Adsett (a large MNC)…eventually they have subsidiaries in overseas countries, including Indonesia…they recommended us to the customer and then the customer contacted us directly by phone… they came to us and said they wanted to install our nursing call system into their hospital…that's how it all began." – (*General Manager and son of the principal founder of an electronics company)*

Here a firm's strong ties with a customer lead the customer to recognize an opportunity for them in the Indonesian market and pass it on.

Opportunity recognition can also come from network members more deliberately combining resources to achieve innovation. For example, a Chinese-German IJV established a dominant position in the Chinese trade show industry with both partners seeking and then combining their own resources. This included the long time managerial expertise of the German partner and their international networks with the Chinese partner's understanding of what would work in their market and their local networks. As well, they combined their capabilities for combining capabilities (based on their separate histories in other ventures) with resulting synergies (Dawson, Young, Tu and Chongyu 2012).

The role of relations and networks in shaping the discovery of new business ideas is reflected in a number of concepts and theories of innovation and technological change, including: Hakansson's theories of industrial technological development from a network views (Håkansson and Johanson, 1992, Hakansson, 1987); the concept of open innovation (Chesbrough, 2003); Von Hippel's concepts of developing links with lead users (Hippel, 1986); and the concept of productive friction (Hagel, 2005).

As pipes and prisms, relations and networks play a key role in the development and exploitation of new ideas and opportunities. They are the means by which the various resources and skills required to commercialise and refine a new idea are accessed and assembled, i.e. networks as pipes, and the means by which such ideas are passed on or sold to others who are better able to exploit them, i.e. networks as prisms (Chandra, 2009, Chandra, 2012, Styles, forthcoming, Wilkinson, 2008).

The networks required to access and assemble relevant resources, skills and capabilities are not necessarily the same as those involved in the creation of the new idea. This is evident in the way countries and firms complain about losing the value of technology inventions developed by them to others. The industrial context required for invention is not necessarily the same as that for commercialization and exploitation.

## 3. Complex Systems Theory: A Theoretical Framework for Understanding Innovation in Business Networks

In order to be better able to understand innovation in and the evolution of business networks and to advise practitioners and policy makers we need a different type of theoretical framework to that usually found in mainstream literature. Mainstream perspectives are dominated by linear, actor focused, comparative static, variables based and reductionist theories. We need to move to nonlinear, network and context oriented, dynamic, process and mechanism-based holistic theories. Such a perspective is reflected in complex systems theory that has its origins in general systems theory that began in the 1950s (Bertalanffy, 1972) and heavily influenced Wroe Alderson's theories of marketing Alderson, one of the founders of modern marketing theory (Alderson, 1965, Alderson and Cox, 1948). We have written about this extensively elsewhere, including the managerial and policy implications, (Wilkinson, forthcoming, Wilkinson and Young, 2002, Wilkinson and Young, 2005, Wilkinson et al., 2012, Wilkinson, 2008) so we will only recap some of the main points here.

As already mentioned, business relations and networks are examples of complex adaptive systems in which order arises in a self-organising bottom-up manner from local actions and interactions taking place. Macro structures and order emerge and are reproduced or not over time in this way in a continuous process of being (existing at a given time and so influencing actions and interactions) and becoming (being reconstituted or changed by the experience and outcomes of the the ongoing actions and interactions taking place over time). Business networks are comprised of the ongoing actions and interactions (activities) of animate actors (people, organisations), inanimate objects (resources, material things, geography) and abstract objects (ideas, schemas and business cultures) operating in an environmental context that its self is a complex systems of other business relations and networks as well as the macro environment – the socio-economic, cultural, biological and material world. These dimensions of business relations and networks are summarized in terms of the actor, activities, resources and schemas that underly much of the thinking of the Industrial Marketing and Purchasing (IMP) Group (Håkansson and Snehota, 1995, Welch and Wilkinson, 2002).

While this theory addresses the complex processes that underpin the evolution of networks, much of the associated empirical investigation does not. Instead there is a growing trend in business to business research to use survey methods and statistical analysis (Denize and Young 2007). Complex systems theory (CST) seeks explanation not in terms of the common statistical variables based models. Variables do not exist in the real world they are abstractions developed by researchers to disembody dimensions of the world from their real world contexts. Instead CST seeks explanation in terms of event sequences, processes and mechanisms, as nicely described by Herbert Simon:

Events unfold over time and are interconnected through various mechanisms and processes. Innovations are events that emerge through time as a result of various processes and mechanisms by which ideas diffuse, mutate, are combined and recombined and evolve. They are part of the ongoing flux of business life. These result in both incremental and substantial changes or innovations arising over time and place. Mechanisms and process refer to why and how events happen. We like to think of them as the "verbs" of explanation and events as the "nouns".

To study innovation in business networks from a complex systems theory we focus on identifying, understanding and modelling the underlying processes and mechanisms in play and how they play out over time within and across firms, relations and networks. This calls for an extended portfolio of research methodologies, moving beyond the commonly-used surveys and experiments, relevant though they still are. This is described in the next section.

## 4. Methodological Implications of a Complex Systems Perspective

To move away from variables based models of innovation we need to directly observe and model the actions of actors innovating and the mechanisms and processes involved. We need to understand what takes place before an innovation occurs that makes it more or less possible, what happens during creative acts of innovation and what happens afterwards that shapes their development, exploitation, success or failure. To do consider the unfolding of this process, this we need longitudinal studies and systematic case histories that map the events taking place in parallel and event sequences and that analyze by which processes events are interconnected. In other words, we need to be able to identify, understand and model the mechanisms and processes involved in linking events over time and place. Lastly, we need to be able to translate this understanding into complex systems computer simulation models that enable us to analyse and explore the role and importance of different mechanisms, processes and environmental conditions on the way innovation happens. Various methodologies are particularly relevant to this endeavour.

### 4.1 Systematic Case Histories

Necessary for investigation of processes are systematic in depth case-histories of past and unfolding innovations including the events leading up to the creative act, the new idea or opportunity, the start of the innovation, and subsequent events including the further development and refinement of the new idea, its exploitation and commercialization (or not) and the consequences arising, such as further innovations, performance outcomes and experience (Buttriss and Wilkinson, 2006, Van de Ven and Engleman, 2004).

The purpose of these case studies is to identify and understand:

- The people, organisations involved and their characteristics, including their aims, resources, skills, orienations, prior knowledge and history as well as the social and business relations and networks they are involved in that are relevant in understanding the causes and consequences of the innovation;
- The main sequences of events leading up to and from the innovation and how they are interconnected;
- The mechanisms and processes at play driving and connecting the sequences of events;
- Relevant contextual effects and key characteristics of the players

Central to this approach are the ways that events can be identified; events are displayed so they can be described and sorted according to pre-identified/theoretically-determined and emergent criteria. Analysis of the description of the history that contains the events and the patterns of the events facilitates identification of connections among them including the mechanisms that drive and connect events.

There are a number examples of such case histories that we have been involved in (Huang, 2010, Buttriss and Wilkinson, 2006, Buttriss, 2009, Welch and Wilkinson, 2005, Bairstow and Young, 2011, Bairstow and Young, in press), One example directly relevant to innovation in networks is research by Bairstow and Young (2011) and Young and Bairstow (2011) who consider the evolution of the Australian IT industry distribution channel over a 20 year period. For the purposes of identification, "events" were defined as markers that were of medium or high relevance in the evolution of the channel with respect to at least one structural condition component (as identified by framework developed from Methlie and Gressard (2006)). Systematic data collection used multiple data sources to gain the broadest possible range of information as to key events and allow triangulation. Primarily this was archival - articles IT trade journals and industry reports. This kind of data avoids problems of imperfect recall and presents reports from many different observers' perspectives. These data were augmented and verified via interviews with eight "experts" (people working in Australian IT distribution for the entire case period).

Systematic analysis of the history involved four steps.. First, archival data was examined sequentially to identify the actors involved, the nature and timing of critical events and the processes taking place were identified and a data base of these was developed. A historical mapping of key events was undertaken. The large patterns emerging (i.e. clusters of actors, events and processes) were identified and as were some of their drivers. The impacts of a major event or series of events were explored and clusters of those leading to a phase change, i.e. a substantive change in the channel configuration that would require equilibration, were identified (Franzosi, 1998, Abell, 1987). This was verified using computer-aided lexicographical analysis (www.leximancer.com).

Four phases in the 20 years of evolution of the channel studied were thus identified and provided a framework for further analysis. The processes that linked the coded events were considered using content and causal analysis. Processes that were highlighted by archival sources (directly and by inference) were compared to the interviews of industry experts with similarities and differences noted. Attributions of importance and causation were noted and triangulated. Frequency of mention by experts and annual review pieces in the trade journal as to key events in the industry for each year were also used to draw conclusions about the importance of events, relationships between them and their impact on evolution. This aggregated material then guided an evaluation as to how and why the Australian IT channel evolved over time. A schematic of the events, phases and identified processes is presented in Figure One.

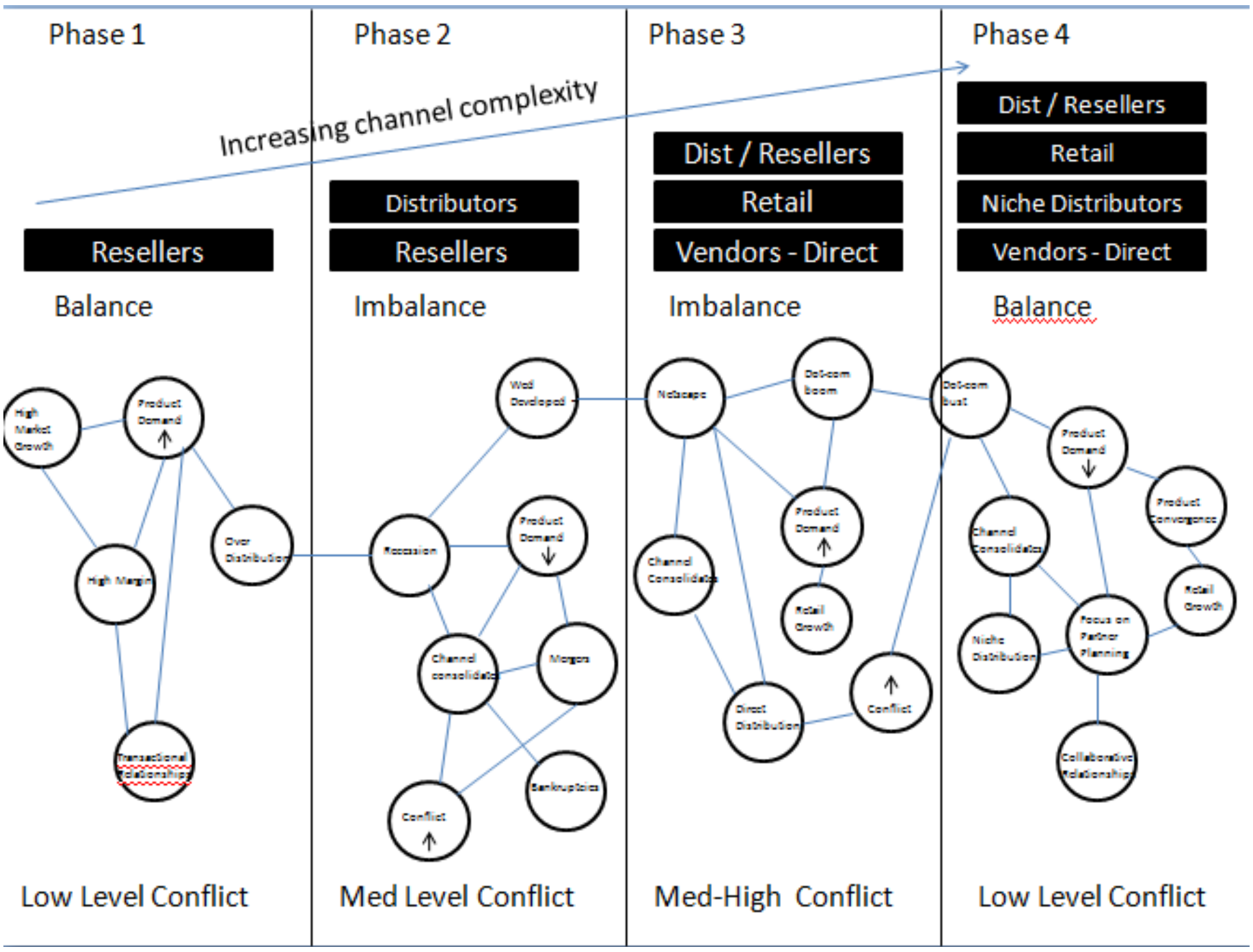

*Figure 1: Processes of Australian IT Channel Change 1986 - 2007 (Source: Young and Bairstow 2011)*

Another useful tool for parsing and connecting the sequence of events is event structure analysis using the Ethno software system (http://www.indiana.edu/~socpsy/ESA/Tutorial.html). This is a programme that assists the researcher or participants in a case to carefully think through each event and identify other prior events on which it depends or is connected to in some way. This has been used to help map and analyse sequences of events in many types of histories, including the development and evolution of innovation through collaboration (Young and Freeman 2008), of business relations and networks (Huang, 2010, Bairstow and Young, 2011) and the evolution of an organization towards becoming an e-business (Buttriss, 2009).

An example of a case history into innovation using Ethno concerns the identification of a new international market opportunity, which may be viewed as a type of innovation or entrepreneurial act (Chandra, 2009, Chandra, 2012). Figure 2 shows a time map of the sequence of events leading up to and from the discovery of an international market opportunity for a pharmaceutical product.

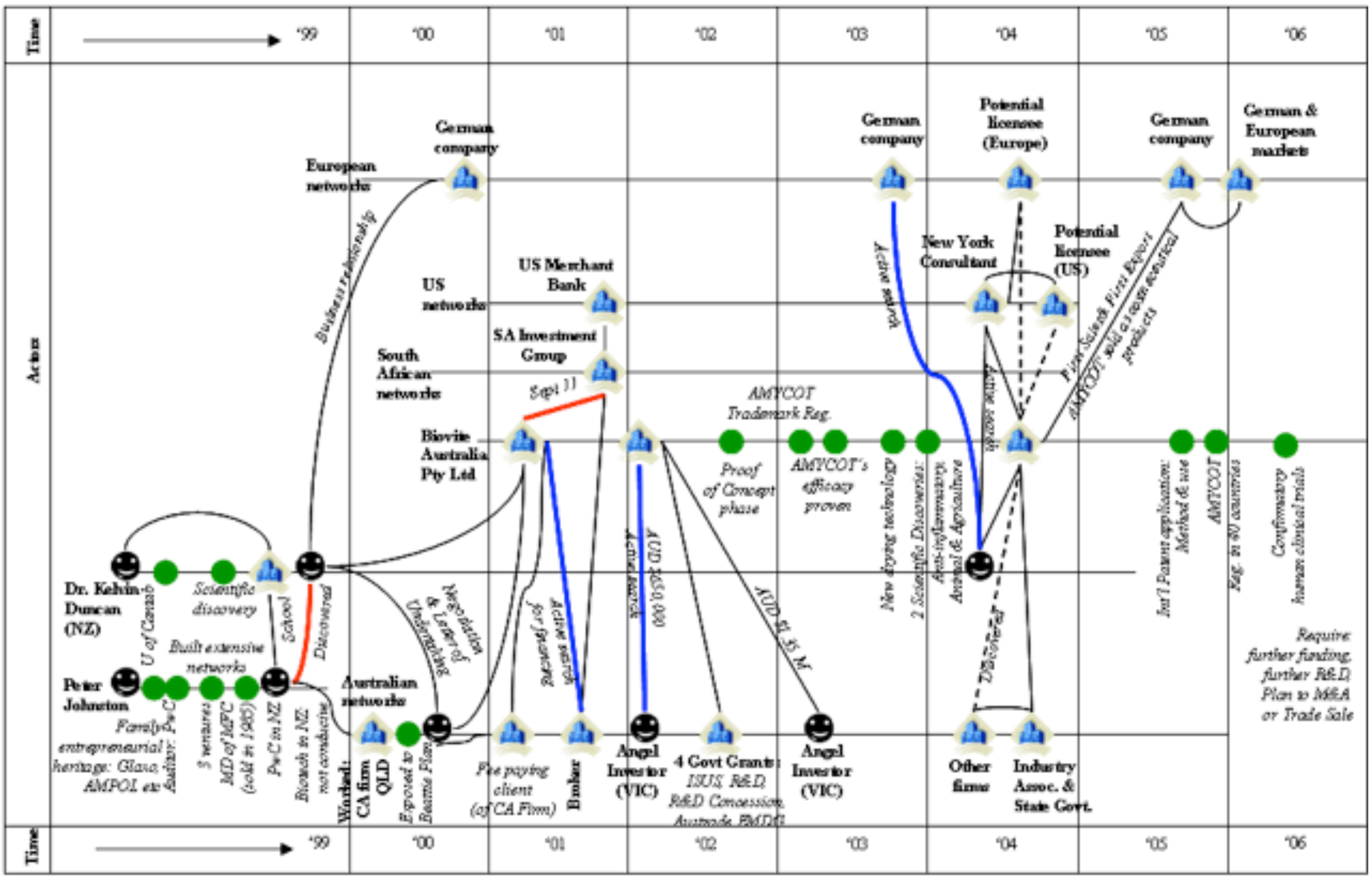

*Figure 2: An Event History Map of an International Market Opportunity Recognition Process (Source: (Chandra, 2012)*

## 4.2. Complex Systems Science

A second methodological advance is to embrace complex systems computer simulation methods and agent based models. Complex systems science is working a quiet revolution in the social sciences (Jorg, 2011, Tesfatsion and Judd, 2006, Epstein, 2006) and there is increasing awareness of it role and value in marketing and business (Rand and Rust, 2011, Wilkinson et al., 2012, Wilkinson, forthcoming). As already noted, business networks are complex adaptive systems. Complex systems simulation methods are concerned with synthesizing as well as analyzing the workings of systems. It seeks to explain by being able to reproduce the essential features of a system, such as a business network, in the form of a computer program. If we are able to reproduce the known behaviour and features of a complex system at the micro and macro levels we have a deeper sense of understanding than one based on measuring and analyzing the behaviour of the real system.

Synthesis has been the source of major advances in science. In the mid 19th Century chemistry advanced when it was able to work out the Periodic Table of the Elements, which showed the building blocks of all known material substance. From this they were able to synthesise new elements and substances that did not exist naturally. This led to the development of many new materials that are an indispensible commonplace of modern life. Some 100 year later biology and biochemistry took a similar gigantic step forward when the structure of DNA was discovered. This revealed the building blocks of life and led to major advances in our understanding of drugs, diseases and life.

In a similar way the ability these days to build nuanced complex systems simulation models rivaling the complexity of the business systems they seek to understand offer the opportunity to advance social and business sciences in major ways. This does not mean we replace existing methods. Far from it. We need these methods to inform, guide, analyse and test our agent-based models. The systematic case histories discussed in the previous section can be augmented by other forms of qualitative enquiry, surveys, natural and lab experiments and the like. Complex systems science complements and extends these existing research methods.

As Chris Langton (Langton, 1996), one of the founders of complex systems research noted, without the ability to synthesise existing and new forms of business life, we are restricted to an effective N of 1 of our world and its history. We are restricted to the studying what has happened, not what could happen.

Agent-based models change this. They are formal mathematical models written in the form of computer code specifying in the form of if-then statements how all the actors in a business network behave and respond (Leombruni and Richiardi, 2005, Borrill and Tesfatsion, 2010). All relevant mechanisms and processes are modeled and calibrated and tested against known features of the real world and environmental and starting conditions are specified. The model can be validated and tested against real systems to ensure it can reproduce known features and the stylized facts of real systems, This can be done at both the micro and macro level (Axtell and Epstein, 1994, Marks, 2007). The technology exists today to build highly complex, nuanced models of vast scale. For example there is a European Union framework project now in pilot stage to build a living earth simulator – a complex systems model of the whole world (www.futurict.eu). Epicast, the Los Alamos National Laboratory Epidemiological Forecasting Simulation Model of the USA, has 300 million agents, one for each person in the USA (www.lanl.gov/orgs/tt/license/software/epicast.shtml)

Innovation processes involve many components processes and mechanisms that can be modeled in various ways. These include models of the diffusion and adoption of information within and across people and organisations, learning and adaptation models, experience curve models and models of entrepreneurship and creativity. Systematic case histories perhaps augmented with other methods will play an important role in identifying and modeling relevant mechanisms and processes driving innovation in business networks and in verifying and validating the worlds those models attempt to simulate.

We believe that ideas are themselves a type of agent in the world and need to be included in our simulation models. This is at the frontier of agent based modeling but seems feasible. Richard Dawkins introduced the concept of memes to refer to ideas that jump from mind to mind and replicate themselves (Dawkins, 1976, Blackmore, 1999). Memes are the cultural equivalent of genes and are subject to similar processes or mechanisms of evolution, i.e. selection, reproduction, recombination and mutation, which have been modeled in various ways (Richerson, 2005).

Once complex systems models have been developed and tested and validated against real world data they can be used in various way by researchers, practitioners and policymakers. They can be realistic experimental laboratories for researchers, who are able to examine the effects of different factors on behaviour and outcomes that would be impossible, unethical and too costly in the real world. They can be flight simulators for managers and policymakers, helping them sharpen their intuitions about the complex behaviour of business networks and explore alternative potential futures.

The tools to do this are becoming ever more powerful, user friendly and freely available. They include sophisticated integrated, Java based, simulation platforms such as RePast (repast.sourceforge.net/) and MASON (cs.gmu.edu/~eclab/projects/mason/), as well as the more easily accessible yet very powerful NetLogo simulation program that was originally designed to teach complex systems programming to primary schools (ccl.northwestern.edu/netlogo/). There are also more training courses and textbooks available these days to learn these methods (Gilbert, 2008, Wilenski, in press, Held, forthcoming). In addition, there are many useful websites to keep you up to date on current development and to provide support materials and demonstration programs, including the comprehensive Agent-based Computational Economics (ACE) site maintained by Leigh Tesfatsion at Iowa State (www2.econ.iastate.edu/tesfatsi/ace.htm). Lastly, there are specialist journals and research conferences that include research on business and markets. The specialist journals include the online *Journal of Artificial Societies and Social Simulation* (jasss.soc.surrey.ac.uk/JASSS.html) and *Advances in Complex Systems* (www.worldscinet.com/acs/). The specialist conferences include the New England Complex Systems Institute and the International Conferences on Complex Systems (www.necsi.org), the European Social Simulation Association Conferences (www.essa2011.org/) and the International Society of Artificial Life Conferences (alife.org/)

## Conclusion

This paper presents an innovative way of thinking about innovation and accompanying this are suggestions for innovative methods of researching it. This two-pronged approach is important. Past research into innovation to some degree has been directed by what was most popular in terms of research methods, which in turn has influenced the way innovation is conceptualized (i.e. we argue that the path between theory and method is two way). Our concerns regarding this in terms of constraining and denigrating business research have been noted elsewhere (Denize and Young 2007). As presented here, new theories about the nature of innovation in conjunction with new methods to research it can substantially progress our knowledge in this area.

So, there is much research to be done and the future looks bright, interesting and fun. Go forth and multiply these these ideas and methods.